ORIGINAL RESEARCH

# Liposomes versus metallic nanostructures: differences in the process of knowledge translation in cancer

David Fajardo-Ortiz[1]
Luis Duran[1]
Laura Moreno[1]
Héctor Ochoa[2]
Víctor M Castaño[1,3,4]

[1]Faculty of Medicine of the National Autonomous University of Mexico, Mexico City, Mexico; [2]El Colegio de la Frontera Sur, San Cristobal de las Casas, Mexico; [3]Molecular Material Department, Applied Physics and Advanced Technology Center, National Autonomous University of Mexico, Juriquilla, Mexico; [4]Advanced Technology Center, CIATEQ, Queretaro, Mexico

Correspondence: Víctor M Castaño
Centro de Física Aplicada y Tecnología Avanzada, Universidad Nacional Autónoma de México, Boulevard Juriquilla 3001, Juriquilla, Querétaro 76230, México
Tel +1 442 211 2657
Fax +1 442 192 6129
Email meneses@unam.mx

**Abstract:** This research maps the knowledge translation process for two different types of nanotechnologies applied to cancer: liposomes and metallic nanostructures (MNs). We performed a structural analysis of citation networks and text mining supported in controlled vocabularies. In the case of liposomes, our results identify subnetworks (invisible colleges) associated with different therapeutic strategies: nanopharmacology, hyperthermia, and gene therapy. Only in the pharmacological strategy was an organized knowledge translation process identified, which, however, is monopolized by the liposomal doxorubicins. In the case of MNs, subnetworks are not differentiated by the type of therapeutic strategy, and the content of the documents is still basic research. Research on MNs is highly focused on developing a combination of molecular imaging and photothermal therapy.

## Introduction

Cancer is a growing global health problem: it is estimated that the incidence of new cases of cancer will double by 2030 compared to 2008.[1] Population aging is the main force that drives the increased incidence of cancer.[2] It has been suggested that the potential adverse effects associated with demographic and epidemiological changes that will occur in the coming decades could be mitigated by technological development.[3] However, the technological innovation process is not guided by the burden of disease except in the case of cancers and cardiovascular diseases, and even when these disease groups are disaggregated, correlation is lost.[4] Moreover, nanotechnology has been mentioned as a promising source of alternative treatments and diagnostic methods for cancer.[5,6] However, until now, there has been no clear evidence that the evolution of knowledge about cancer nanotechnology will lead to innovations in the fight against cancer. On the other hand, it is not well understood how the disruptive character of cancer nanotechnology[7] could affect its translation into clinical applications.

A technology is disruptive because it breaks with the "normal" line of technological development of a particular class of products.[8] Disruptive technologies allow the entry of new competitors in a given market, which implies a potential threat to incumbents.[9] The opposition between dominant and disruptive technologies is relative: there is a technology development cycle in which dominant designs were originally disruptive innovations.[10] In this study, we propose that nanotechnologies are not a homogeneous group and that they are in different stages of evolution of technological development.

**2627**

In this study, we compared the knowledge translation for two different kinds of nanotechnologies – liposomes and metallic nanostructures – which, because of their temporality, are clearly in two different stages of technological development. The former is still in a preclinical stage, whereas the latter is already a reality in the pharmaceutical market. Our objective is mainly methodological and exploratory: we are developing a way to map the knowledge translation through scientific literature networks in health nanotechnologies; along that line, we are using these two very different types of nanotechnologies as case studies, since this allows us to obtain a contrasting view of how different the structure, organization, and translation of knowledge look in regard to a consolidated technology, such as liposomes, and for a very new therapeutic tool, such as metallic nanostructures.

Liposomes are "phospholipid bubbles with a membrane bilayer structure",[11] which, because of their size, are located on the border between micro- and nanotechnology.[12] Liposomes were among the first nanotechnologies to be used in clinical trials and among the first to appear on the market.[13] Regarding their applicability, Sen and Mandal point out that:

> Efficient entrapment of therapeutics, biocompatibility, bio-degradability, low systemic toxicity, low immunogenicity and ability to bypass multidrug resistance mechanisms has made liposomes a versatile drug/gene delivery system in cancer chemotherapy.[14]

Liposomes are the dominant design among nanotechnologies applied to cancer treatment.

The biomedical use of metallic nanostructures, mainly gold, has been identified as a new paradigm of cancer treatment and diagnosis.[15,16] Gold nanostructures have the advantage of being nontoxic, inert, easy to synthesize, and versatile.[17] However, the most revolutionary qualities of metallic nanostructures in the development of cancer treatments are their optical properties. Surface plasmon resonance of metallic nanostructures increases the absorption and scattering of light, which has application in cancer imaging, spectroscopic detection, and photothermal therapy.[18,19] The metal nanostructures are a radical innovation compared with liposomes.

As mentioned above, the current study seeks to compare knowledge translation in these two nanotechnologies applied to cancer. The knowledge translation has been studied in citation networks through two methodological strategies. The first strategy is to classify documents into two areas of research: the first area is focused on the "discovery", while the second is focused on the "delivery". Subsequently, network analysis is performed to identify the "main paths" in each field, and, finally, to identify citations that connect them.[20,21] The second strategy is to classify journals into four categories ranging from basic research to clinical observation, according to the terms used in the titles of the articles, and then map the citations among the journals.[22,23]

An alternative methodological approach to the two mentioned above was developed by us to study the translation of knowledge on cervical cancer.[24] This strategy is based on the modeling of networks of highly cited papers, which are semantically analyzed using controlled vocabularies of the Medical Subject Headings (MeSH)[25] and the Gene Ontology (GO).[26] This strategy, by and large, was the one used in this investigation.

## Methods

The following steps were undertaken in this study:

1. A search for articles (November 2013) on cancer and liposomes, or cancer and metallic nanostructures, was performed in the Web of Science (WOS).[27] The search criteria are shown in Table 1.

2. The 20% most cited articles, which together accounted for at least 60% of the citations in their respective fields, were selected and downloaded from WOS.

3. Cytoscape version 2.8.1 (The Cytoscape Consortium, San Diego, CA, USA),[28] HistCite[29] (Eugene Garfield, Thomson Reuters Corporation), and Pajek version 3.14[30] (Vladimir Batagelj and Andrej Mrvar, University of Ljubljana) software were used to construct two network models. The first network model corresponds to cancer research and liposomes, while the second comprises documents on cancer and metallic nanoparticles.

4. The scientific papers that constitute the body of network models were sought on the GoPubMed[31] website (http://www.gopubmed.org/web/gopubmed/), which semantically analyzes and labels the documents with GO and MeSH terms.

5. The Clust & See[32] application (Laboratoire TAGC/INSERM U1090, Marseille, France) was used to identify subnetworks (subnets) within the models. Identification of subnets and distribution of MeSH terms was conducted to identify research areas within each field.

6. MeSH terms that are embedded within the higher hierarchy categories "Diagnosis", "Therapeutics", "Surgical Procedures, Operative", "Named Groups", and "Health Care" are considered clinical terms. The terms that are outside of these categories, and GO terms are considered nonclinical terms.

7. The proportion of clinical versus nonclinical terms was calculated for each document in the network model. This ratio is a measure of how clinical or basic a paper is.

8. Each document that is part of the network model was tagged with information about the institution, city, and country of origin.

## Results

### Liposomes

Following the criteria outlined in Table 1, 1,456 papers on cancer and liposomes were found. Of these, 291 (20%) were selected. These were cited 22,949 times, representing 68.2% of all citations (33,657) found in the WOS. The proportion of citations shows the importance of these documents in the process of scientific communication.

One hundred and fifty-two of the selected articles form a single network of citations. By analyzing the network model using the Clust & See software, nine subnets were identified (Figure 1). The distribution of GO and MeSH terms identified for each subnet reveals that several of these correspond to different lines of research (Figure 1). Subnet 1 is related to the development of liposomes as vehicles for drugs in general; subnet 2 is related to the clinical use of liposome-encapsulated doxorubicin; subnet 3 is related to basic research into doxorubicin liposomes; subnet 4 relates to clinical research of gene therapy; subnet 5 also relates to gene therapy but at the level of basic research; subnet 7 is associated with hyperthermia therapy; and subnet 9 is related to small RNA interference.

Research on liposomes and cancer is basic except in subnet 2 (Figure 1), which is composed mainly of Phase II and Phase III clinical trials and multicenter randomized trials. Research on liposomes and cancer is largely dominated by the United States, followed by Europe, Canada, Japan, and Taiwan (Figure 1). The University of Texas MD Anderson Cancer Center, Houston, TX, USA stands out as the leading institution, with 15 papers included in the network model (Table 2).

## Metallic nanostructures

Six hundred and seventy-seven articles on metal nanostructures and cancer were found, of which 137 (20%) were selected, which were cited 12,878 times, representing 76.3% (16,884) of all citations made to the documents.

Eighty-four of the selected nodes form a single network of citations. By analyzing the network model using the Clust & See software, six subnets were identified (Figure 2). However, the distribution of GO and MeSH terms for each subnet shows no differences between subnets in terms of clinical or biomedical content, although differences in the combination of technology do exist. Subnet 1 is characterized by the combined use of metal nanoparticles with carbon nanotubes, while, in subnet 2, the research is characterized by combining laser and nanoparticles. Finally, subnet 6 relates to phototherapy. Subnets 3, 4, and 5 are associated with terms that are common to all documents in the network model. Generally, the items that make up the network model are basic research, with some scattered clinical research articles in subnets 1 and 2 (Figure 2).

Research in cancer and metallic nanostructures is strongly dominated by the United States, followed by China and South Korea. The institution leadership is split between the Rice University in Houston, TX, USA and the Georgia Institute of Technology in Atlanta, GA, USA, with seven documents for each institution within the network model.

## Discussion

Liposomes and metallic nanostructures differ in their network structure, knowledge translation, and leadership by country and institution. How relevant are these differences and how are they related to the concepts of disruptive technologies and dominant designs? What can these maps tell us about the forces that guide the development of these nanotechnologies?

**Table 1** Search criteria of the articles in the Web of Science

| Nanotechnology | Title | Topic | Document type | Timespan |
|---|---|---|---|---|
| Liposomes | Cancer* or carcinoma* | liposome* | Article | All years |
| Metallic nanostructures | Cancer* or carcinoma* | "gold nanoparticle*" or "au nanoparticle*" or "metal nanoparticle*" or "silver nanoparticle*" or "ag nanoparticle*" or "gold nanorod*" or "au nanorod*" or "metal nanorod*" or "silver nanorod*" or "ag nanorod*" or "gold nanoshell*" or "au nanoshell*" or "metal nanoshell*" or "silver nanoshell*" or "ag nanoshell*" | Article | All years |

**Note:** *wild card.

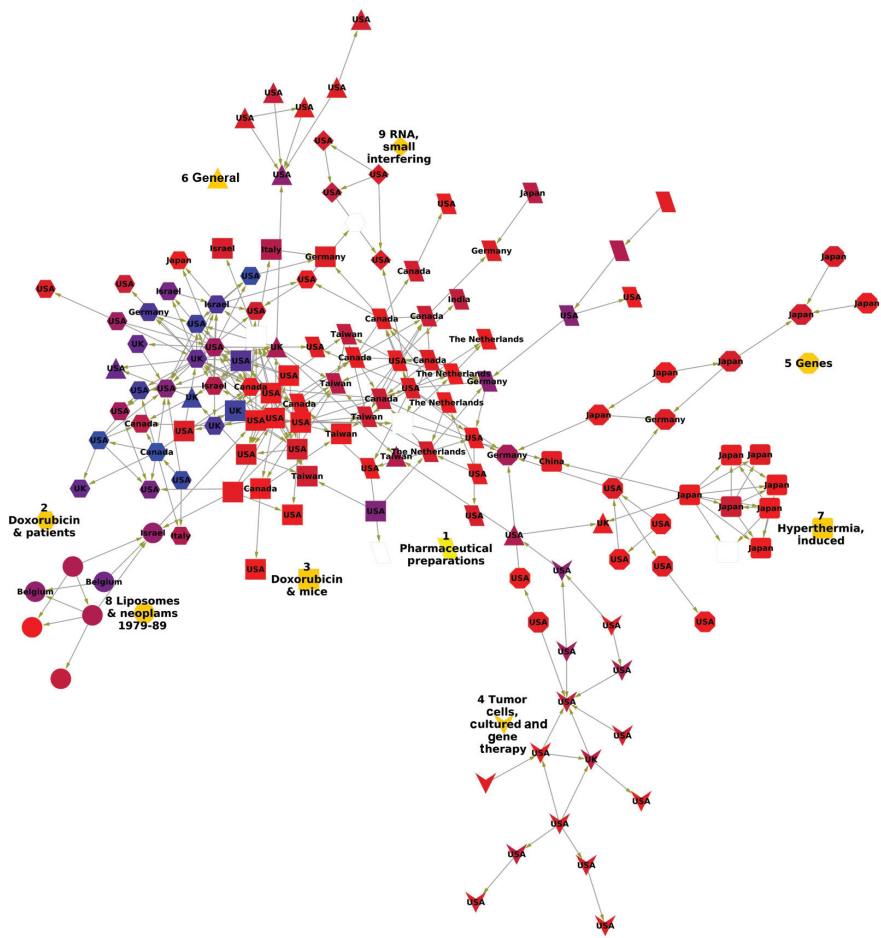

**Figure 1** Network model of research papers on cancer and liposomes.
**Notes:** Each node represents one paper of the 20% most cited papers on liposome research applied to cancer, and the edges represent the citations between the documents (nodes). The shape of the nodes indicates to which subnetwork they belong (1–9). The color of the nodes is according to a continuous scale from red to blue. This scale is a function of the clinical terms rate, so a red node could be considered a basic research paper, a purple one clinical research, and a blue node is a clinical observation article. Nodes without text are papers that did not report any address.

Let us begin by briefly discussing the scope of the results. The maps (Figures 1 and 3) show that the most cited papers cite each other. High citation is not necessarily related to the quality or validity of a research paper;[33,34] however, the most cited papers, which in turn cite each other, represent the paradigmatic core of a given area of knowledge.[35] If the contents of these documents reflect the essential components of a particular paradigm, analyzing the content is relevant since paradigms dictate the guidelines on how to understand and treat a health problem.[36]

## Liposomes

The network model shows how research on liposomes and cancer is dominated by a pharmaceutical approach over other potential therapeutic strategies such as hyperthermia and gene therapy. This pharmaceutical approach is in turn dominated by the use of liposomes as carriers of doxorubicin. The results also show a translation of basic research into clinical knowledge through the interaction between the two subnets associated with doxorubicin (subnets 2 and 3; see Figure 1), which is related to the fact that Doxil® (Janssen Products, LP, Horsham, PA, USA) was the first nanodrug approved by the US Food and Drug Administration (FDA) in 1995.[37]

Our research was not focused on the line of innovation and development of Doxil, even though this may be in relative terms the dominant design of liposomes applied to cancer, because our search form focused on identifying the "paradigmatic body" of the liposomes and cancer research, with emphasis on documents whose headings include the words "cancer" or "carcinoma". However, FDA approval of Doxil was a milestone that likely radically affected the evolution of knowledge on liposomes and cancer. Twenty-six of the 28 documents that belong to subnet 3, which relates to clinical research on liposomes, are subsequent to 1995, and 12 of them are about Doxil. These documents dominate the communication process in subnet 3 (Figure 3).

**Table 2** Leading institutions in liposome and cancer research

| Institution | Location | Number of papers |
|---|---|---|
| The University of Texas MD Anderson Cancer Center | Houston, TX, USA | 15 |
| University of Alberta | Edmonton, AB, Canada | 8 |
| Nagoya University | Nagoya, Japan | 7 |
| Roswell Park Cancer Institute | Buffalo, NY, USA | 6 |
| University of California, San Francisco | San Francisco, CA, USA | 6 |

The oldest and most central document in subnet 3 is about Doxil (Figure 3) is a Phase II clinical trial aimed at treating advanced breast cancer.[38] Given that Doxil was originally approved for the treatment of Kaposi's sarcoma,[37] we conjecture that subnet 3 corresponds to the phase of imitation of Doxil previously reported by Venditto and Szoka,[39] whose document is based on a model of inventors, innovators, and imitators inspired by the ideas of Schumpeter.[40] The invention phase is made up of scientific and technological developments prior to the innovation stage, which begin with the optimization of remote loading of the drug in liposomes, continue with clinical trials, and end with approval by the FDA,[39] while the imitation phase seeks to improve the product or extend its use to solve other problems. As such, subnet 3 would focus on extending and/or improving the clinical use of Doxil. This makes sense if we consider that this methodology identifies the paradigmatic body of normal science in Kuhnian terms, not in revolutionary science that is equivalent to a phase of technological innovation.

One question arose constantly throughout this investigation: why is there such a marked hierarchy, in which Doxil dominates over other liposomal doxorubicin formulations and the latter dominate the pharmaceutical approach, which in turn dominates research on liposomes applied to cancer treatment? In other words: why is there so little diversity of anticancer products using liposomes?

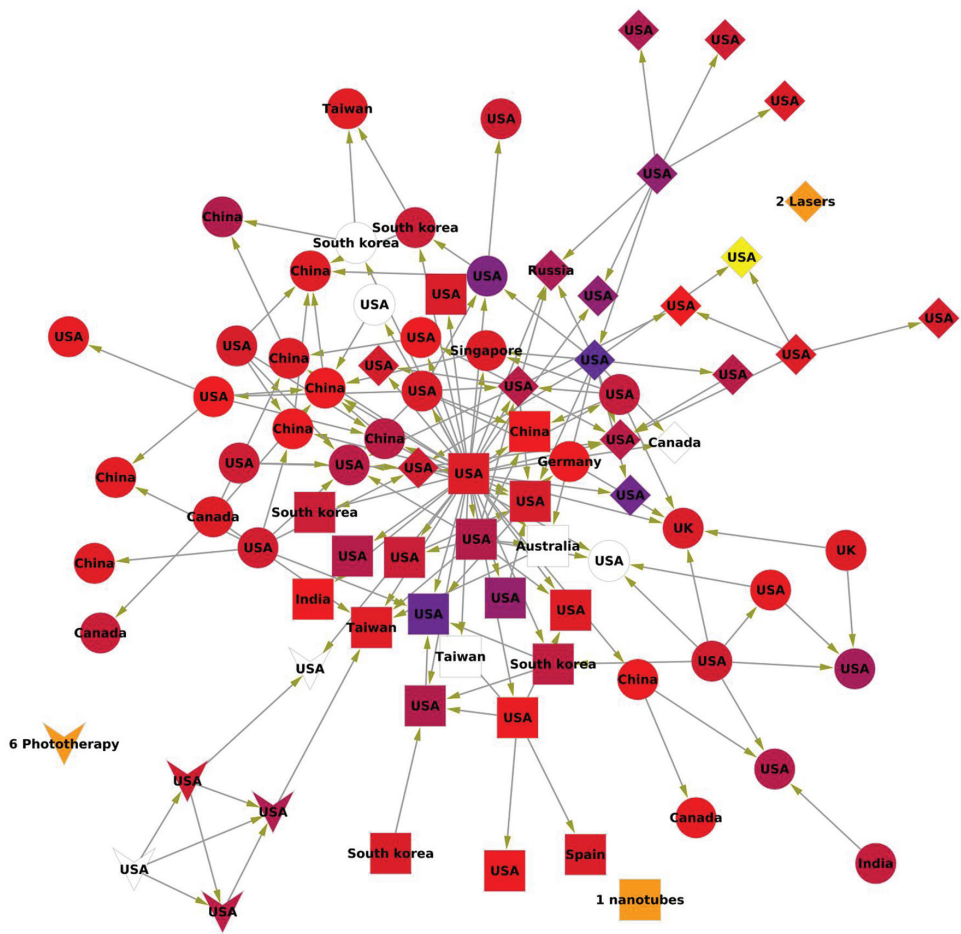

**Figure 2** Network model of research papers on cancer and metallic nanostructures.
**Notes:** Each node represents one paper of the 20% most cited papers on liposome research applied to cancer, and the edges represent the citations between the documents (nodes). The shape of the nodes indicates to which subnetwork they belong. The color of the nodes is according to a continuous scale from red to blue. This scale is a function of the clinical terms rate, so a red node could be considered a basic research paper, a purple one clinical research, and a blue node is a clinical observation article. Subnet 1 is related to nanotubes, subnet 2 is related to lasers and subnet 6 is related to phototherapy; the other networks (3–5) are not displayed because they are generals.

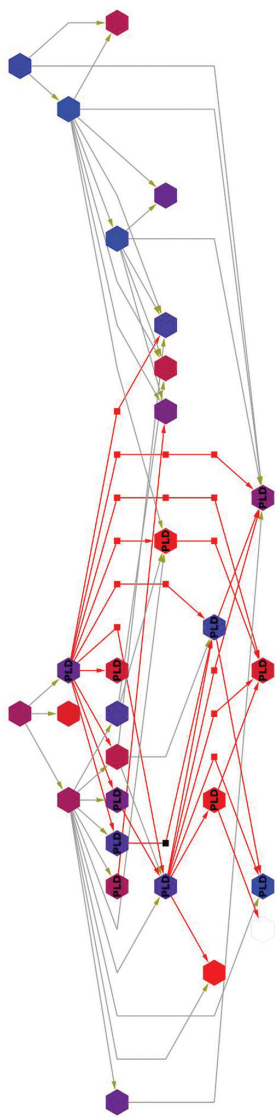

**Figure 3** Subnetwork 2 hierarchically organized.
**Notes:** Each node represents one paper of the 20% most cited papers on liposome research applied to cancer, and the edges represent the citations between the documents (nodes). The color of the nodes is according to a continuous scale from red to blue. This scale is a function of the clinical terms rate, so a red node could be considered a basic research paper, a purple one clinical research, and a blue node is a clinical observation article.
**Abbreviation:** PDL, pegylated liposomal doxorubicin.

Three clues can lead us to a partial answer to this question:

1. Maine et al[41] identify three strategies that led to the invention of Doxil: 1) importation of ideas from a variety of fields of knowledge, 2) the creation of an atmosphere of deep collaboration, and 3) the fit between technology and market. We could reinterpret the above as a complex process of negotiation between different fields of knowledge and actors, and between market and technology, all of which are deeply related to the concept of trading zones.[42] The emergence of these areas requires, inter alia, the construction of a new creole language.[43] This requires dialogue and negotiation between fields of knowledge and actors which carry with them a set of prior epistemic commitments, several of which must be broken.

2. Barenholz, one of the inventors of Doxil, describes the technical difficulty of developing this nanodrug due to a combination of physicochemical constraints. For example, the liposomes need to be nanosized in order to be favored by the effect of enhanced permeability and retention, but decreased size affects drug loading into liposomes.[37] In the same paper, Barenholz notes the difficulty for imitators to develop a generic version of Doxil that succeeds in being approved by the FDA,[37] which highlights the weight that the combination of regulatory and technical barriers could have when trying to innovate with nanodrugs.

3. A third argument is related to the difference in time of approval between conventional drugs and nanomedicines. Of this, Venditto and Szoka said:

   This is probably due to the fact that for most currently approved drugs, reformulating them in a nanocarrier provides a small increase in performance that large pharmaceutical companies do not consider being worth the time, effort and expense of development.[39]

As pointed out in the second paragraph of the discussion, our results are focused on the paradigmatic literature, leaving out the most recent inventions within the field of liposomes applied to cancer. One of these promising recent inventions is the folate-targeted liposomal (FTL) zoledronic acid (ZOL). It has been reported that dipalmitoyl phosphatidylglycerol-FTL-ZOL showed better in vitro results than PEGylated FTL-ZOL, non-folate targeted liposomal ZOL, free ZOL, and doxorubicin;[44] however results in in vivo preclinical research indicate that liposomal ZOL was severely toxic for the animals (mice).[45] Another research group reported positive therapeutic results in a mouse xenograft model with increased survival, antitumor activity, and tolerability,[46] with better results obtained using PEGylated calcium/phosphate nanoparticles as ZOL carriers.[47] A future study comparing the structure and knowledge translation between different drug nanocarriers could be relevant.

## Metallic nanostructures

Our results indicate that the subnets are not distinguished by the therapeutic approach or way of understanding cancer, which makes sense given the disruptive, novel (the documents that make up the network were published between 2004 and 2012), and basic character of the documents in the network. The content of these documents involves a first interdisciplinary dialogue between the biomedical sciences and metallic nanostructures. The central documents of subnets 1 and 2 illustrate the general strategy of this particular type of nanotechnology, which involves taking advantage of the increased absorption and scattering of light, and the rapid conversion of light into heat energy that occurs in metallic nanostructures in order to develop molecular-scale imaging and photothermal therapy.[48,49] This strategy represents a technological breakthrough when compared with liposomes, the liposomes found a niche in the pharmaceutical and biotechnology industry. Metal nanostructures have not yet left the universities and research centers; as such, it is difficult to define a path that tells us how this nanotechnology will reach the patient. The differences in the leading institutions and countries where these two different kinds of nanotechnologies are developed further emphasize the disruptive nature of the metal nanostructures. Institutional leadership in the case of liposomes lies with institutions that are global leaders in terms of research and treatment of cancer, while, for the case of metallic nanostructures, leadership lies in the institutions that are already leaders in nanotechnology. Asia replaced Western Europe as the United States' partner in the development of metallic nanostructures applied to cancer. The last statement is consistent with observations that suggest that China is fast becoming a leading nation in nanotechnology on par with the United States,[50] and that nanotechnology can change the landscape of research and development on a global scale, allowing the emergence of new technological powers.[51]

## Conclusion

This research is the first attempt to map and compare the structure and translation of knowledge about specific types of nanotechnologies applied to cancer. Liposomes are a disruptive technology that is in the process of becoming a dominant design, exhibiting a process of knowledge translation and branches that correspond to different therapeutic strategies. However, there is a marked dominance of the pharmaceutical approach, which is organized around liposomal doxorubicin. While liposomes are moving from a stage of innovation to imitation, metallic nanostructures are still in a phase of invention, in a first interdisciplinary dialogue between the biomedical sciences and nanotechnology, which revolves around the development of photothermal therapy and molecular imaging. Finally, the translation of this research strategy into a useful tool for decision makers is currently under way and will reported separately.

## Author contributions

All authors contributed to the research design, interpretation of results, and writing of this paper. David Fajardo-Ortiz built the database and performed the network analysis and text mining.

## Acknowledgments

David Fajardo-Ortiz is supported by a CONACYT PhD scholarship. The Digital Medical Library of the Faculty of Medicine, UNAM, provided access to the Web of Science, which enabled this research.

## Disclosure

The authors report no conflicts of interest in this work.